\documentclass{svjour2}                

\smartqed

\usepackage{latexsym}
\usepackage{amsmath}
\usepackage{amsfonts}
\usepackage{amssymb}
\usepackage{graphicx}

\newcommand{\chs}[1]{{\left\{#1\right\}}}
\newcommand{\prs}[1]{{\left(#1\right)}}
\newcommand{\prob}[1]{{\mathcal{P}\left(#1\right)}}
\newcommand{\ket}[1]{{\left|#1\right>}}

\newcommand{\cut}[1]{}

\journalname{}
\begin{document}

\title{A Bayesian Foundation for Physical Theories \thanks{This project was partially supported
by EPSRC grant EP/E049516/1}}

\author{Roberto C. Alamino}

\institute{Roberto C. Alamino \at
           Non-linearity and Complexity Research Group, Aston University\\
           \email{alaminrc@aston.ac.uk}}

\date{Received: date / Accepted: date}

\maketitle

\begin{abstract}
Bayesian probability theory is used as a framework to develop a formalism for the scientific method based on principles
of inductive reasoning. The formalism allows for precise definitions of the key concepts in theories of physics and also
leads to a well-defined procedure to select \emph{one or more} theories among a family of (well-defined) candidates by
ranking
them according to their posterior probability distributions, which result from Bayes's theorem by incorporating to an
initial prior the information extracted from a dataset, ultimately defined by experimental evidence. Examples with 
different levels of complexity are given and three main applications to basic cosmological questions are analysed: (i) typicality of human observers, (ii) the multiverse hypothesis and, extremely briefly, some few observations
about (iii) the anthropic principle. Finally, it is demonstrated that this formulation can address problems that were out of the scope of scientific research until now by 
presenting the \emph{isolated worlds} problem and its resolution via the presented framework.
\end{abstract}

\section{Introduction}
\label{section:Introduction}

Science is fundamentally based on the processing of information accumulated as we humans continually observe the
endless spectrum of natural phenomena occurring in our universe. The philosophical foundations of the task of trying
to make sense of all this collected information in a rational way, which received the name of \emph{scientific method}, were constructed based on the principle that \emph{scientific knowledge} should rely entirely on the empirical evidence
collected from these observed phenomena and that the theories elaborated to \emph{explain} them should not make any 
unnecessary assumption beyond what is required to account for the facts. Constructing a scientific theory would then
be equivalent to inferring patterns, or an underlying connecting structure, from this acquired information. 
The very hypothesis of
this possibility relies on the also empirical observation of the regularity and predictability of nature.

The above paragraph is full of ill-defined concepts, but they contain the essence of what we would require from an 
\emph{ideal}
scientific theory. One of our aims in this work will be to give more precise definitions not only to the above presented ideas, 
but also to many others that appear in connection with physics in particular and with science in general. It is 
important to clarify at this point what the theory to be developed here \emph{is not}. The present work \emph{is not} a 
sociological theory of how scientists develop the scientific knowledge. It is not concerned with historical details of 
how theories were discovered or developed and how were the mental processes that resulted in their final
formulations. A theory like that would have to take into consideration not only sociological and psychological aspects,
but also their complex relations with the economical and political environment in which the theories appeared. 
The theory, although a better word would be \emph{framework}, to be developed here will be concerned with what could
be considered the \emph{ideal} mathematical methods that should be used by the scientific inquiry and, if not, 
will lead to wrong or suboptimal
results. The author thinks that he does not have the proper training to analyse the historical contingencies of the 
development of theories by humans and that, in fact, social scientists and historians are better suited to this task.

That said, let us point out that physics can be considered the best example of the application of the scientific method,
providing us with
descriptions of nature with astonishing \emph{quantitative} accuracy as in the well-known case of quantum electrodynamics. The success of
physics is indisputable as can be easily attested by the improvement of quality of life brought by our present
technology. This high efficiency of physics is a direct result of taking seriously the basic tenets of the scientific
method to discern between what should be considered science and what should not. 

One of the most fundamental of these criteria is the concept of \emph{falsifiability}, which is a central point in the
Popperian description of science \cite{Popper59}. Falsifiability can be a difficult concept to identify in theoretical
physics, specially when the theoretical advancements run ahead of the experimental technology, as is the present situation with respect for instance to quantum gravity, where the difficulty of falsifying tentative theories is well recognised \cite{Amelino08}. Partly because of that, physicists rely also on other characteristics of physical theories, like mathematical simplicity and elegance, as guides in the development of theories even if they are not
rigorously defined concepts. Obviously it must be always stressed that these criteria are only guidelines that must be
discarded if experiments turn out to disprove them. A rather illustrative example, although \emph{apparently} nonsensical  
for today scientists, is the idea that every substance is composed of a mixture of only four elements: water, fire, air and earth. This can be considered a rather elegant theory, as it requires only four basic elements in comparison with
the more than one hundred elements known today, but nonetheless it was proved wrong beyond doubt by the overwhelming
experimental evidence accumulated since its formulation.    

Given that what tells science apart from mysticism and other non-scientific human activities is the strict reliance on
rational thought, many attempts were made to formalise the scientific method using inductive logic, with deductive
logic as a particular case (see text and references in \cite{Gower96}). Salmon \cite{Salmon71} is a good example of one of first authors to analyse concepts involved in inductive logic applied to scientific explanations with the introduction of his idea of \emph{statistical relevance}. Although since the very early stages 
it was recognised
that inductive, probabilistic reasoning should underpin these efforts, it is fair to say that the correct formulation
was only achieved with the modern development of Bayesian inference \cite{Jaynes03}. The work of Salmon, for instance, can be neatly mapped to Bayesian language with appropriate care as he himself suggests in the above cited book. The application of Bayesian inference to the problem of formalisation of the scientific method
however has not yet made use of the full power of the theory, although there exists at least one very developed
tentative framework, called Minimum Description Length (MDL) \cite{Li97}. 

MDL fuses concepts of information theory, computation and probability theory in an attempt to develop a method to
decide between theories given some data. The main idea of MDL is to choose the best hypothesis by calculating the 
length of the smallest description in a certain code simultaneously of the hypothesis and the data when compressed using the hypothesis itself. We will not give any detailed account of MDL here, but a comparison of its main points with the
work we develop here will be made in section \ref{section:Conclusions}. 

It can be shown that Bayesian inference is the only way to deal with information that comply with a sensible set of
requirements for inductive and logical reasoning \cite{Jaynes03,Caticha07,Caticha08}. Although many misconceptions
about the Bayesian interpretation of probability are still used against it on fundamental grounds, with strong attacks
against its use and the use of probabilistic reasoning in general within the scientific method \cite{Brody94}, the
success of Bayesian methods in machine learning applications, which are ultimately an attempt to understand and
formalise an ideal process of thinking, is an experimental fact \cite{Sivia06,Engel01}. These fundamental questions
will not be discussed here and the interested reader can refer to Jaynes' book \cite{Jaynes03} for a thorough study of
these and other fundamental points. A huge supporting bibliography containing all the relevant works can also be found
there and, in order to save space, will not be presented here.

In this work, we will first use the Bayesian inference framework to introduce our reasoning that will lead to a 
formalisation of a concept of \emph{scientific method} concept, which is done in section \ref{section:Bayes}. The basic processes that compose the scientific methodology according to this framework, which will be called \emph{information acquisition}, \emph{modeling}
and \emph{testing/selection}, are there introduced and briefly commented. The precise meaning of information acquisition is explained in section \ref{section:Data}. Section \ref{section:Modeling} then analyses the modeling process of theories in a more detailed way.
Testing, which we will use as a short word for testing/selection, is studied in section \ref{section:Testing}, after which follows a discussion about the important related concept of 
falsifiability in section \ref{section:Falsifiability}. In particular, this latter section will contain the two most 
important definitions of this paper, those of a \emph{scientific theory} and of the \emph{scientific method}. These sections contain the core of our framework and the next sections will be concerned with applications. In section \ref{section:Cosmology} the formalism developed here
is used to analyse three current cosmological issues in theoretical physics, namely the question of typicality of human
observers, the preference for a multiverse description and a brief analysis of the meaning of the (weak) anthropic principle. Section \ref{section:Problem} will introduce the \emph{isolated worlds} problem and show how the present framework yields a solution that extends the applicability of scientific method beyond today's range. A discussion and the final conclusions are given in section \ref{section:Conclusions}.

\section{Bayesian Reasoning and Physical Theories}
\label{section:Bayes}

Usually, most discussions about the scientific method begin with a philosophical consideration
of \emph{truth}. The concept of truth can be stated in more pragmatic terms by asserting
that there is a well defined underlying process generating every acquired piece of information from nature. 
This is not done here as the formalism to be developed does not require this concept to be
considered at all, even if a completely positivist point of view is taken.

Every formalisation of a subject starts by trying to establish precise definitions of the 
key concepts the theory wants to capture. In the case of the scientific method, there are many intuitive and 
philosophical requirements that scientific theories should 
fulfill. A non-comprehensive list includes Ockham's razor, falsifiability, elegance and explanatory power. Some of these 
concepts are easier to formalise than others. We will try to address those which seem most important in our opinion.

Before that, however, let us first consider a simplified model to begin our discussions. The most fertile ground for this is machine learning. The basic task of this discipline is to study mathematical models, usually to be implemented in computers, that are able to do some kind of ``learning''. Learning is mainly associated to identifying patterns in a
given dataset. Consider as an example the \emph{perceptron}, a very simple machine learning model thoroughly studied by
methods of statistical physics \cite{Rosenblatt62,Engel01}. 
A perceptron is a mathematical model where a \emph{student perceptron}
characterised by a vector $S\in\mathbb{R}^M$ tries to infer a rule given a set of examples. In a 
\emph{supervised learning} 
situation, the rule is encoded by the vector $B\in\mathbb{R}^M$, which is usually called the \emph{teacher perceptron}. 
The student then has access to a dataset $D_t$ formed by $t$ pairs $\prs{\xi_i,\sigma_i}$, $i=1,...,t$ 
of \emph{questions} $\xi_i$, which are usually $n$-dimensional binary vectors, and \emph{answers} $\sigma_i=f\prs{B,\xi_i}$, usually taken to be Boolean function, and has to infer from it the value of $B$. 
The function $f$ is called 
the \emph{activation function} of the teacher and in the simplest scenario is
known to
the student, although we can consider a situation where the student will have to infer its
functional form as well \cite{Neirotti03}. This process of ``discovering the
rule" is called \emph{generalisation} and has as a measure the so-called \emph{generalisation error} which is the 
probability of the student to answer wrongly a new question. Note that the
activation function can have an stochastic component, which is usually associated with some
kind of noise that distorts the dataset. 

The perceptron was inspired by the landmarking paper of McCulloch and Pitts \cite{McCulloch43} which introduced a 
simplified mathematical model for real human neurons and initiated the field
of artificial neural networks \cite{Hertz91}. In fact, \emph{neural networks} are formed by
collections of interconnected perceptrons.
Similarly, but in a more complicated way, when humans 
do science the main objective is to try to infer natural rules, which are supposed to exist, based on some dataset encoding information derived from observations about a system. Although nature may be independent of humans, science is realised by human brains, which are devices that evolved to process information, making this processing an important part of science.
The process of human thinking by deductive logic has been studied since ancient times with a theory of deductive logic
being already proposed by Aristotle.
However, pure deductive logic is not enough to capture all characteristics of thinking. It can
be shown \cite{Jaynes03,Caticha08} that given some set of reasonable requirements that 
plausible reasoning must follow in the form of \emph{Cox axioms} \cite{Cox46}, probability theory is the 
correct mathematical description of it. In addition, Bayesian inference becomes the correct
method to update probabilities based on new evidence. As it is natural to require that the scientific method follows the 
formal rules of plausible reasoning, Bayesian inference is the framework we choose to formalise it.

The concepts to be formalised in the following have been used in many recent discussions 
about fundamental aspects of physical theories, specially cosmological models \cite{Page07,Hartle07}. A precise 
definition of these concepts may help not only to better
pose the relevant questions, but also to make sure that all conclusions are correctly derived
from the analysis of known data and possible hypothesis.

The aim of the scientific method is to generate scientific theories. Intuitively, a physical
theory $T$, or any other scientific theory actually, is constructed to ``explain" some observed
facts, although the meaning of ``explaining" is a philosophical question which will not be
explored in depth in this work. We will assume that what we call \emph{observed facts} can either be described as the 
result of physical measurements obtained from experiments chosen
from some set, or the information obtained by some kind of reasoning, as the result of a mathematical theorem that is suppose to answer some mathematical question. The experiments or questions in general will be assumed to be chosen from a \emph{question set} $\Xi$, to be better defined in the next section. In addition, we also assume that the results obtained as answers to the question, the result of a measurement for instance, can always be taken as real vectors, as any information can in some way be encoded by them, although sometimes a non-numerical description of them 
may be more illustrative, as in the case where the measurement corresponds to some quality as ``colour" for instance. This 
process by which information is acquired is a basic part of the scientific method and we will call it 
\emph{information acquisition}. Acquiring information is both required to construct and to test a theory.
Using the same notation we used for the perceptron, $D_t$ will denote a collection of $t$ data
points, with each data point composed by a label denoting the question and the corresponding acquired answer.

The process of constructing a theory, or \emph{modeling} will be discussed in details in 
section \ref{section:Modeling}. With the knowledge of some dataset $D_t$ it is possible to 
elaborate many different theories about the acquired information. When faced with different theories
then, we would like to be able to compare their \emph{predictions} to some dataset in
order to assess their relative agreement with the information we were able to extract. 
Consider, for instance, a set of $M$ theories $T_1,...,T_M$ created to
explain a dataset with questions drawn from some set $\Xi$. According to the the discussion up to this point, the most appropriate method to choose the ``best" theory according to a set of first principles (which we assumed to be \emph{Cox axioms}) is to rank them by their \emph{posterior probabilities} given the data using Bayes' theorem 
\begin{equation}
  \label{equation:Rank}
  \prob{T_i|D_t} = \frac{\prob{D_t|T_i}\prob{T_i}}
                   {\sum_i \prob{D_t|T_i}\prob{T_i}}.
\end{equation}  

The probability $\prob{D_t|T_i}$ of the observed dataset given the theory, also known as the likelihood of the
theory, represents the idea of ``predictions" of the theories. This probability is normalised for 
fixed values of $t$ as
\begin{equation}
  \sum_{D_t} \prob{D_t|T_i} =1.
\end{equation}

Of course it is possible to include a probability distribution over the value of $t$ as well. Suppose, for instance, that the probability of taking a measurement of some sort is constant in time. Then, the probability of taking $t$ measurements in a fixed interval of time is given by a Poissonian distribution
\begin{equation}
  \prob{t} = \frac{\bar{t}^t e^{-\bar{t}}}{t!},
\end{equation}
where $\bar{t}$ is the average value of the number of measurements. However, this can be easily
accommodated in the formalism by including $t$ as a random variable in all the probability formulas and we will avoid this sophistication as it does not affect any of the main considerations.

The probability distribution $\prob{T_i}$ in equation (\ref{equation:Rank}) is called the \emph{prior
distribution} of the theories. It gives a ranking of preference of the theories according to
those intrinsic characteristics of each theory which are independent of the dataset. Note that by fixing the number of
theories to be compared ($M$ according to the given notation) we avoid any ambiguities with respect to the fact that we never really know how many theories are possible. We will always be concerned with comparing theories that we have, not ones that we still do not. 

Equation (\ref{equation:Rank}) is the fundamental formula of Bayesian inference and will also be our
central principle. Based on the analysis of its consequences, we will be able to study the
process of testing physical theories, which will be dealt with in details in section \ref{section:Testing}. 

It may be argued that we should be able to evaluate the plausibility of a theory even if we
have only one. But then, by using equation (\ref{equation:Rank}) we would always find probability
one for this theory as there is not any other. That is not a weakness of the procedure, for
it is designed to choose one between many possible theories. To evaluate the probability of one
theory being right according to the data, the correct procedure is to consider a secondary hypothesis: is
the theory correct or wrong? Then, these two hypothesis can be compared using Bayesian
inference giving the desired result. To a more detailed discussion, the reader is referred to
\cite{Jaynes03}. This latter case however does not weight other characteristics of a theory
and would not discriminate between two different theories that however explain the data with the same 
precision.

Theories cannot be proved right, but can be proved wrong, even if only in principle.
We are only keen to consider a theory
scientific if we are able to falsify it, which is accomplished by making predictions that 
can be checked by comparing with the results of the experiments. This procedure is part
of the testing process. Still, falsifiability is such an important concept that we will
dedicate the entire section \ref{section:Falsifiability} to its study.

As it can be appreciated from the exposition above, information acquisition, modeling and testing are interrelated, dependent processes. In fact, we claim that these three basic elements are
all that is needed to formalise the scientific method. Science should then be carried out
by iterating the above processes in whatever order it is necessary to produce a probability 
distribution for scientific theories at some instant $t$ after processing all the information available in the 
dataset $D_t$. 

Note that this does not produce a unique theory unless the probability of all other theories is reduced to zero at some 
point. Selection of one theory, if desired, can be carried out by 
choosing the most probable theory after some dataset has been measured. In the next sections 
we will start to develop the formalism and give more precise definitions that will enable us to study in details each one of the processes described above.

\section{Information Acquisition}
\label{section:Data}

In order to be able to formalise the scientific method, we will first attempt to give definitions for its basic concepts. It became clear by the arguments of the last section that a fundamental concept of the theory should be that of a \emph{question} and we therefore will give tentative formal definitions to the basic elements that are required for it as well as try to relate them to their natural (physical, mathematical) interpretations.

\begin{definition}[Questions and Measured Answers]
  A \textbf{question} $Q\in\mathcal{Q}$ is a mathematically describable object encoding all the information about a
  specific physical experiment or mathematical reasoning. A \textbf{measured answer} $A\in\mathcal{A}$ to a 
  question $Q$ is another describable object that encodes the information acquired by carrying out the
  experiment/reasoning encoded by $Q$, which is represented symbolically by
  \begin{equation}
    A=\rhd Q.
  \end{equation}
  The sets $\mathcal{Q}$ and $\mathcal{A}$ can be of any kind of describable objects, not necessarily the same.  
\end{definition}

The concept of a \emph{question} encompass both the entire concept of a physical experiments and that of a mathematical/logical reasonings. For a physical experiment, this must encode the experimental setup, the method of measurement and any other relevant information that is necessary to reproduce it. In the case of a reasoning, the method
of reasoning and the hypothesis used to obtain the measured answer should be part of it. The encoded information should also includes initial conditions related to the systems under study or characteristics of these systems like, for instance, the mass or the charge of a particle in a high energy experiment.
 
The freedom allowed with respect to the nature of questions and measured answers is necessary to guarantee that any
kind of inquiry is possible. For instance, in the case of the perceptron, the questions are multidimensional euclidean vectors with coordinates usually being boolean or real \emph{and} the activation function. On the other hand, questions can also be formulated in more familiar languages, like for example ``What is the mass of the electron?'' or ``Are there any integers $x,y,z$ satisfying $x^n+y^n=z^n$ for integer $n>2$?''. In quantum mechanics, as another example, questions can be associated with the calculation of eigenvalues of Hermitian operators (and any mathematical method used to do it) with measured answers corresponding to their actual numerical values. 

\begin{definition}[Question Set]
  Consider an index set $I$. The set
  \begin{equation}
    \Xi=\chs{Q_i}_{i\in I},
  \end{equation}
  where each element $Q_i$ indexed by values of the index set $I$ corresponds to a possible
  question about a system, is called a \textbf{question set}. 
\end{definition}

The index set $I$ can be either discrete or continuous. It also can be finite or infinite. We
will not restrict it in any way a priori. Note that the index of the questions can be related to numerical parameters 
encoded by it. This has the profound consequence that the index 
set $I$ can then, in some situations, acquire a physical interpretation beyond that of a simple mathematical object.
Indeed, we know that concepts as mass and charge, for instance, label in quantum field theory
the possible representations of symmetry groups and correspond to the characteristics that define the elementary particles.

\begin{definition}[Dataset] Given a question set $\Xi$ as defined above, the set
  \begin{equation}
    D_t =\chs{(\xi_\mu,\sigma_\mu)}_{\mu=1,...,t},
  \end{equation}
  of ordered pairs of questions $\xi_\mu\in\Xi$ and corresponding measured answers $\sigma_\mu=\rhd\xi_\mu$, called
  \textbf{data points}, where $t$ is an integer, is called a \textbf{dataset}.     
\end{definition}

The notation is borrowed from machine learning (see again the perceptron example given in the previous section). 
Although we are using the notation $t$ to index the dataset, this must not be interpreted as a time index in general,
but only in particular cases if it is designed as such. We could try to generalise $t$ by allowing its values to be in
any set, but this is unnecessary in most practical situations and therefore we will not attempt to do that. 

There are many cases where the experiments $\xi_\mu$ can be chosen independently of each 
other, but it may be conceivable that this does not happen in many situations. For
instance, suppose the acquired information corresponds to values of some Markovian random variable at
each time step. Then it is possible to choose the initial state of the system, but the 
following states will be defined by its stochastic dynamics and are not independently chosen.
The specific way into which the correlations are encoded is
defined by the theory $T$, which will be defined in the next section, and is part of its structure.

It is important to notice at this point that in the case of physical experiments, the resulting value can be corrupted
by many different kinds of noise. Usually, the \emph{central limit theorem} guarantees that when many sources of noise
are acting together, the final result is a Gaussian noise, but this may not always be the case. In the same way as
the dataset correlations, we will see that the concept of noise and its modeling also depends on the theory $T$ being
considered and, as such, will be described in more details when we discuss the process of \emph{modeling}, in the next
section. As a trivial example, electric experiments are made with devices with a noise model that depends on electromagnetism itself.

At many points during the scientific investigation of a physical process, experiments are made in an attempt to obtain 
more information to continue the process of modeling. In these cases, there are many methods that can be used to
decide about what kind of questions should be asked. An
interesting approach is named \emph{learning by queries}. Queries are questions that are chosen in a certain way as to
maximise the amount of information acquired with their answers. The theory of learning by queries has been analysed and
used successfully in many applications in neural networks with the possibility of being generalised to other situations.
We will not delve into details here, more information being available in \cite{Sollich94} and references therein. 

Let us then finish this section by given the definition of the first of the three processes that will compose the
scientific method.

\begin{definition}[Information Acquisition]
  The process of constructing a dataset $D_t$ by obtaining the measured answers to some subset of a the question
  set $\Xi$ is called \textbf{information acquisition}.
\end{definition}

\section{Modeling}
\label{section:Modeling}

The process of modeling is the most involved aspect of the scientific method. Here is where
the inspiration of the researcher comes into play. There is no algorithmic methodology at present to create a model, only guidelines. A formalisation of this process would require traits
like creativity to be modeled and, until we have a better understanding of the process of
thinking, this will remain uncertain ground. Modeling the process of creating a theory is not
our objective. Here it is assumed that there is a way to model the theory and we will try
to assess its characteristics. We start by giving a more precise definition of a theory. Note that we still are not defining a \emph{scientific} theory, for which we still need some more considerations.

\begin{definition}[Theory]
A \textbf{theory} $T=(\alpha,\pi)$ is composed by an algorithm $\alpha$ depending on a set of $p$ free parameters 
$\pi=\chs{\pi_1,...,\pi_p}$, called the theory's \textbf{fundamental constants}, that generates in a finite time a probability distribution $\prob{D_t|T}$ for any possible dataset $D_t$ 
with questions belonging to a specific question set $\Xi$. In this sense, we say that the theory $T$ \emph{answers} the question set $\Xi$ and the probability distributions $\prob{D_t|T}$ are called the \textbf{theoretical answers} of the
theory.
\end{definition}

In physical theories, the fundamental constants are called \emph{physical constants}, which must be
dimensionless \cite{Duff04}, and this is the motivation for the adopted terminology. The values of the theory's
constants are not part of the theory and need to be obtained from experiments, or better, from the measured answers.
The correct way to obtain them, once more, is to use Bayesian inference to extract its values from the available
datasets. We will talk more about information processing using the the datasets in section \ref{section:Testing} when
we discuss how to test the theories. 

In principle, it can be argued that the algorithm $\alpha$ can be either stochastic or deterministic. This is, however,
just a matter of convention. A stochasticity of an algorithm $\alpha$ can always be transferred to its theoretical answers by modifying the resulting probability distributions of the datasets accordingly. However, there is a sense in which we can talk about a deterministic or stochastic theory that is reminiscent of physical theories. This can be formalised in the following way.

\begin{definition}[Stochastic and Deterministic Theories]
  A \textbf{deterministic theory} is one where \emph{all} theoretical answers are delta functions, i.e., probability
  distributions with zero variance. Otherwise, the theory is called \emph{stochastic}. 
\end{definition}  

It is worthwhile to clarify at this point that we are not trying to delve into the philosophical debates about the 
meaning of the mathematical structure of the theory. This can be illustrated by the question of the interpretation of 
quantum mechanics. The mathematical structure of quantum mechanics is well defined and it is this structure that we are
calling the \emph{theory}. The pictorial representations in terms of human language or perception, although being 
extremely important in our opinion, are questions that are out of the scope of our framework and will not be dealt with
here beyond what we consider to be necessary to develop our ideas. However, we agree that this is an interesting 
future research direction.

The given definition of a theory is minimal in the sense that the internal structure of the algorithm is
not completely specified. We can however try to identify this structure. Consider a dataset $D_t$. As already pointed
out in the previous section, the theory must be able to describe correlations between its data points, the noise affecting the experiments (in the relevant situations) and the probability distributions corresponding to ideal
theoretical answers in the absence of noise.

The interdependence of a certain set of variables can always be represented by a \emph{Bayesian network} 
\cite{Pearl88}, which is a \emph{directed acyclic graph} where each node represents one 
of the variables and each directed edge represents conditional dependence. A Bayesian network is a particular case of 
more general probabilistic structures known as \emph{graphical models} which have been used in applications of
statistical mechanics for some time now. Most well known statistical mechanical models, like the Ising model, can be written in this language. By using the information contained in the network and the chain rule of probability,
$\prob{D_t|T}$ can be broken down into factors representing the correlation structure in the dataset according to
the theory. An interesting feature is that, in addition to the nodes representing the data 
points, a Bayesian network may contain \emph{hidden nodes}, which are quantities that
cannot be directly observed but their state can influence the other variables, being able to
encode non-explicit correlations. We will see how these hidden nodes appear in a physical
context when we analyse an example of a quantum mechanical system later on. 

\begin{figure}
  \centering
  \includegraphics[width=5cm]{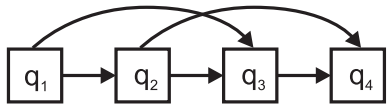}
	\caption{A Bayesian network representing four variables $q_1,q_2,q_3,q_4$ and its dependences. }
	\label{figure:BN}
\end{figure}

Figure \ref{figure:BN} shows an example of a Bayesian network representing a Markov model of order two involving
four variables $q_1$, $q_2$, $q_3$ and $q_4$. Following the dependence encoded in the graph we can then write
\begin{equation}
  \prob{q_1,q_2,q_3,q_4} = \prob{q_4|q_3,q_2}\prob{q_3|q_2,q_1}\prob{q_2|q_1}\prob{q_1}.
\end{equation}

Note that when the dataset comes from a physical process developing in time, the case where the conditional
probabilities of the system states at \emph{every} instant given the former instants are delta functions, represents a 
process where there is no random component in the time development, which conforms to our definition of a \emph{deterministic theory}. The classic comparison is between Newtonian mechanics, which in a fundamental level does not include
any statistical element, versus quantum mechanics, which is fundamentally based on stochastic processes. Even with the
time development of the wave function being still \emph{deterministic}, the results of what we will see posteriorly to be completely noiseless measurements, the ideal theoretical answers of the theory, remain stochastic and are given by probability distributions of possible values which are not delta functions. 

Let us now focus on the characterisation in the theory of how noise affects the answers. The modeling of the noise
presupposes that the experimental conditions encoded by the questions can affect the answers in such a 
way that what would be some unique value in the absence of noise becomes a probability distribution over values. Therefore, this ``corruption'' can be described by some probability distribution. This is easier seen in
the case of only one data point $(\xi,\sigma)$. The generalisation is then straightforward. 
Let us then assume that $\sigma=\rhd\xi\in\mathcal{A}$ is the corrupted value obtained as a measured answer. 
Let us call the \emph{noiseless value} of this answer by $a$, which will not be \emph{experimentally} accessible due to the noise. According to Bayesian inference theory, we should then marginalise over all unmeasurable possibilities and
write
\begin{equation}
  \begin{split}
    \prob{\xi,\sigma|T} &= \sum_a \prob{\xi,\sigma,a|T}\\
                        &= \sum_a \prob{\sigma|a,\xi,T}\prob{a|\xi,T}\prob{\xi|T}.
  \end{split}
\end{equation}

If the way the questions (or experiments, for instance) are chosen do not depend on the theory, the last term in this
formula cancels out when the posterior distribution for the theory is normalised.
The term $\prob{a|\xi,T}$ is then a distribution of possible answers of the question $\xi$ (or experimental results)
without the interference of the noise that we then call the \emph{noiseless theoretical answer}. The noise modeling is contained then in the first term $\prob{\sigma|a,\xi,T}$.
Of course, if there is more than one data point, a Bayesian network specifying the correlation of all variables,
including the possible measured answers $\rhd (\cdot)$ should be given by the theory. 

The algorithm $\alpha$ should then describe all the above stated relationships. It is known that algorithms can be described in many ways by using different languages. The algorithm $\alpha$ also can be seen as a mathematical structure defined by a set of $a$ axioms $\chs{\alpha_1,...,\alpha_a}$ dependent on the fundamental constants $\pi$. A
method to encode mathematical structures developed in model theory is explained for instance in \cite{Tegmark07}.
The physical experiments encoded in the question set $\Xi$ and the possible values of measurements, the measured 
answers, are then associated with theorems derived from these axioms. 

The construction of the algorithm $\alpha$ is usually subjected to many constraints. The most fundamental one being 
(mathematical) consistency. For example, if a physical theory can calculate, let us say, the entropy of a black hole by 
several different methods, they must give the same answer. Other constraints nevertheless can be imposed like,
for instance, Tegmark's Computable Universe Hypothesis \cite{Tegmark07}, which could be
implemented by requiring that the probability distributions which the theory is supposed to
generate must be \emph{computable}, i.e., the theory must allow for their calculation in a
finite time, which in particular we have already included in our definition of theory. Any constraint to be enforced should ultimately be part of $\alpha$ and be incorporated in its axioms in some form. It is then fair to think about
both the amount of axioms needed to construct the theory and the number of free parameters in it as leading in some
way the concept of \emph{complexity} of the theory.

The concept of complexity is obviously not a simple one. There are many measures of complexity in the literature 
and new ones are proposed from time to time, with the exact idea of what is the meaning of ``complex'' being different
in each of them.  A popular measure that seems to be able to capture many desired characteristics is the one
introduced by Solomonoff \cite{Solomonoff60} in 1960 and independently five years later by Kolmogorov
\cite{Kolmogorov65}. It is known as \emph{algorithmic complexity} or alternatively as \emph{Kolmogorov complexity}
(KC). The KC of a given object is defined as the length of the shortest description of an algorithm that can 
reproduce the object in some universal language \cite{Li97,Cover91}. Although this length changes with the particular
language used to express the algorithm, Kolmogorov was able to show that the descriptions differ only by an additive
constant. KC is however an accurate concept only for dealing with algorithms of classical information theory. If we
broaden our spectrum of theories by allowing $\alpha$ to be not only a classical algorithm, but also a quantum one 
\cite{Nielsen00} then the usual KC is not an appropriate measure of complexity in general. 
There are for this case proposals for considering a quantum version of KC \cite{Berthiaume01,Vitanyi00}, 
however there is no clear general complexity measure that is able to encompass both cases. 
This problem is related to the fact that the frontier between the classical and the quantum is not completely 
understood yet.

Even if we restrict our analysis to classical algorithms, Kolmogorov complexity has still one 
drawback, namely, the fact that it is uncomputable. One solution proposed in the MDL
approach is to use instead the prefix Kolmogorov complexity (PKC) which is the restriction of
Kolmogorov complexity to self-delimiting codes, for which there always exists a Turing Machine
that can identify if a codeword is or is not part of the code in finite time \cite{Li97}. 

Be it classical or quantum, the basic principle behind Kolmogorov complexity is that the
complexity of any object is actually related to its regularity, or more specifically,
to its \emph{compressibility}. Consider a classical string of zeros and ones. The more regular
the string is, the smaller is the program needed to reproduce that string. In other words,
we can compress the string in a number of characters less than its size. In fact,
compressibility is related to the number of symmetries of an object $O$ and is fundamentally
linked to the size of its automorphism group $\mbox{Aut}(O)$, or the group of its symmetries.
The more symmetric an object is, the shorter the description needed to reproduce it. The
relationship between $\mbox{Aut}(T)$ and complexity of description for a theory $T$ is 
discussed also in \cite{Tegmark07} and is behind the idea that a GUT should have a larger 
symmetry than the effective low-energy theories unified by it. 

The exact measure of complexity to be used will not be important for our discussion, only the fact that it is a 
quantity that is possible to measure and can approximate the relative importance of this characteristic in each 
theory. The reason being that we would like to compare the relative complexity in order to use it as an additional
criterion for selecting theories which will be included in the prior distribution $\prob{T}$. In fact, there may
be situations where KC or PKC may not be the more convenient choices and simpler 
approximations adapted to the theories being analysed would capture the relative importance of
either theory better than these options. 

There is a subtle point here. Note that in the above considerations about complexity we only talked about algorithms.
The way we defined it, a theory is obviously not only its algorithmic part $\alpha$ and complexity is a concept that should be extended to include somehow the dependence of the theory on the set $\pi$ of its fundamental constants as well.  However,
contrary to the approach taken in MDL, we \emph{choose} to consider the complexity of these two components of the
theory  separately for reasons that will be clearer as we proceed in our study. We still associate the word 
``complexity'' with some measure of the length of the algorithm $\alpha$, but we deal with $\pi$ in a different way. 
It is important to note that what is considered to be a fundamental constant in one theory can be a quantity that may depend on other fundamental constants in a different theory and, therefore, may be calculable in it. 
This latter theory would then have fewer fundamental 
constants then the former and intuitively it would make sense to say that it is ``more fundamental''. This suggests a
way to define a concept which we will call the \emph{fundamental status} of a theory related to the set of its
fundamental constants $\pi$. 

\begin{definition}[Fundamental Status]
  Given two theories $T_1$ and $T_2$ answering the same question set $\Xi$ with sets
  of fundamental constants given respectively by $\pi^{(1)}$ and $\pi^{(2)}$, theory  $T_1$ 
  is said to be \textbf{more fundamental}, or to possess a higher \textbf{fundamental status}, than theory $T_2$ if 
  the cardinality of $\pi^{(1)}$ is
  less than that of $\pi^{(2)}$, or stated in another terms, if theory $T_1$ has less
  fundamental constants than theory $T_2$. 
\end{definition} 

If two theories answer the same question set, even if it is a very restricted one, it is reasonable to expect that their fundamental status and complexity should be related. In fact, as the more fundamental theory will have
less constants, the decrease in the number of constants must result in an increase in the number of relations between
the remaining ones, which implies in an increase in the complexity of $\alpha$. In other words, a theory gets rid of
constants by including constraints that relate its values. From this simple observation we then are led to the
startling idea that \emph{the more fundamental the theory, the more complex it is}. At first sight this must
seem highly counter-intuitive. This obviously is a result of the highly difficult task of defining \emph{complexity} in 
a manner that agree with all the intuitive ideas that we usually associate with this word. However, upon closer
analysis, all concepts involved are intuitively reasonable when we think that theories that we would usually classify
as highly fundamental, consider the example of string theory for instance, rely on less physical constants but are
mathematically much more challenging, to the point of taking the efforts of many physicists during a considerable
amount of time to provide ways to use it to calculate quantities that would be simpler in theories consider less
fundamental. This is not an isolated example. We can easily realise that theories we call more fundamental in general, require much more subtle concepts and much more mathematical sophistication than the others. Actually this could be
understood as the relation between generalisation and fitting in machine learning \cite{Bishop06}. Less fundamental
theories favour fitting as they have more adjustable parameters, while more fundamental ones favour generalisation. We
will have more to talk about that in the next section.

Again, the above expressed ideas seem to be at odds with our assertion that GUTs should be more
symmetric theories as symmetries allows for compression of data. In fact, although symmetries
allow for the reproducibility of an object by giving less information about it, the description
of the symmetry should now be incorporated to the description of the program that reproduces the object, 
which means that although the input that generates the object is now smaller, the program that interprets it should be larger because it must ``know'' somehow what the symmetry is and how to implement it.

A very important observation must be made at this point. At the most fundamental level, it is possible to encode the
algorithm $\alpha$ plus the set of parameters $\pi$ with some assigned values into a number $\mathcal{N}$. This number 
can be fed to a universal Turing machine $\Upsilon$ that then can provide the required probability distributions of the
data points. In this sense, there is only one theory $\Upsilon$ with \emph{only one} numerical parameter $\mathcal{N}$
that is then chosen based on the dataset by using Bayesian inference. This description blurs the
difference between parameters and algorithm that we defined. However, this is a question of hierarchy. We indeed want
to differentiate between a set of axioms and numerical parameters, even if they can be considered the same thing at a
higher, purely mathematical, level. We then can break down $\mathcal{N}$ in independent free parameters and view axioms
as fixing some of these free parameters which then become what we called $\alpha$. These fixed parameters are not
allowed to change even if more information is acquired. Changing them is equivalent to consider a different theory. The
remaining free parameters are allowed to change as more information is collected without considering the result a
different theory. Although we can see by this that the precise concepts of a theory and its constants is a question of
attributing semantic ``baggage'', as considered in \cite{Tegmark07}, this separation is important for relating the input of this theoretical Turing machine $\Upsilon$ to the physical world or whatever system is being studied.

In the following paragraphs we will analyse some basic examples that will serve to illustrate many of the concepts 
discussed above. The examples given are only intended to clarify the definitions introduced, applications of the theory 
being left to the appropriate sections.

\begin{example}[Fundamental Status and Complexity]
  Let us illustrate our basic idea that the more fundamental is a theory, the more complex it tends to become with an
  example of a very simple system system, a set of analytical real valued functions with real arguments. In this  
  set, every function can be expanded in a possibly infinite polynomial around zero. Let us consider a sequence of
  theories where the question set is composed by the possible values of the their argument and the answers are real
  numbers corresponding to applying the function we are trying to discover to these values. A set of five possible
  theories is given by
  \begin{align}
    T_1 &= \text{The function is a polynomial},\nonumber\\
    T_2 &= \text{The function is a polynomial of finite degree $n$},\nonumber\\
    T_3 &= \text{The function is a polynomial of degree 3},\nonumber\\
    T_4 &= \text{The function is a polynomial of degree 2},\nonumber\\
    T_5 &= \text{The function is $\sin x$}.\nonumber
  \end{align}
  
  Note that in $T_1$ there is an infinite number of coefficients to adjust from the data, but the theory is very simple
  to state. In fact, we could say that it has only one axiom. $T_2$ have much less free parameters, as now the number
  is finite. By \emph{adding one extra axiom} saying that the polynomial has a finite degree, we reduced drastically the
  number of free parameters and now we have a more fundamental theory. 
  
  It could be argued that $T_1$ should be more fundamental than $T_2$ as the latter is a special case of the former. 
  However, this is a misleading argument, as being very general, $T_1$ in fact leaves much more structure to be
  adjusted by data than $T_2$. Note that the most general theory is the one that actually assumes nothing, as any other
  theory can be obtained from it by ``adjusting'' its infinite number of free parameters.
     
  Considering again $T_2$, we see that in fact we could break down the finiteness axiom of its degree into a
  \emph{countably infinite} number of axioms by saying that we are fixing an infinite number of parameters, although
  the exact number of non-zero parameters is now a new parameter. $T_3$ and $T_4$ now have one more axiom that fixes
  the degree $n$. The number of free parameters is again reduced, increasing the fundamental status of the corresponding
  theory in the process. Intuitively, it would be hard to decide which one of $T_3$ and $T_4$ is more complex, but 
  accordingly to our conjecture that a more fundamental theory should be more complex, then $T_4$ should somehow be
  attributed higher complexity, although it is not clear for us at the moment how to do that in a rigorous way.
 
  Finally, the theory $T_5$ is obviously more complex than all the others as it requires the specification of
  \emph{all} the polynomial coefficients with the recursion formula that defines the sine function. In contrast, 
  it is the more fundamental theory of all once it simply has \emph{no adjustable parameter}.
\end{example}

\begin{example}[Perceptron]
  Let us see how the concepts developed here are related to the perceptron, already described 
  briefly in section \ref{section:Bayes}. For simplicity, let us suppose that the possible
  question set is $\Xi=\chs{0,1}^n$, for some integer $n$. In fact, it becomes 
  convenient to enumerate the questions using as an index the question itself. For instance, 
  if we had $n=2$ we would write the question set as
  \begin{equation}
    \Xi=\chs{Q_{00},Q_{01},Q_{10},Q_{11}}=\chs{00,01,10,11}.
  \end{equation}
  
  Suppose the activation function of the teacher perceptron $f(B,\xi_i)$ to be a Boolean
  function, by which we mean that its range is in the set $\chs{0,1}$. 
  In physical terms, the teacher perceptron is a ``toy universe" representing nature and its 
  activation function is what we would call the ``physical laws" of the model. The vector $B$
  plays the role of a physical constant, in fact, the only physical (theory's) constant existent in this
  ``universe". The questions can be seen as binary strings encoding every possible experimental
  setups in this universe that are designed to measure one binary digit as a result.  
  
  Let us assume that we start our modeling with a certain dataset of $t$ measurements
  \begin{equation}
    D_t=\chs{(\xi_1,\sigma_1),...,(\xi_t,\sigma_t)},
  \end{equation}
  with $\xi_i\in\Xi$ and $\sigma_i\in\chs{0,1}$, from which we suppose 
  (in this case correctly, but this is not always true) that the ranges of
  the possible answers are all equal to $\mathcal{A}_i=\chs{0,1}$.
  Again, by a stroke of genius or a lucky guess, we suppose that physical laws depend only
  on one physical constant, namely $S$, which in perceptron terminology is the student's
  \emph{synaptic vector}, through some boolean function $g(S,\xi)$, the activation function of 
  the student. Then our physical theory, call it $T^p$, will have a set of physical
  constants which is the singleton
  \begin{equation}
    \pi^{(p)}=\chs{\pi_1^{(p)}}=\chs{S}.
  \end{equation}

  First, let us decide about the dependence of the data points. Our theory will consider
  data points are independent and, being so, we can write their probability distribution
  as
  \begin{equation}
    \begin{split}
      \prob{D_t|T^p} &= \prod_{i=1}^t \prob{\xi_i,\sigma_i|T^p}\\
                     &= \prod_{i=1}^t \prob{\sigma_i|\xi_i,T^p}\prob{\xi_i|T^p}.
    \end{split}
  \end{equation}
  
  In the majority of applications, the probability distribution of the questions given by the
  last term $ \prob{\xi_i|T^p}$ does not depend on the theory and consequently
  this term is canceled by the corresponding term in the normalization of equation 
  (\ref{equation:Rank}). The remaining factor describes how the actual obtained answers are related to
  the the possible questions according to the theory. This represents in general the noise model, but we will
  assume that noise is negligible in our experimental setup. The probability for our given
  dataset is then modeled by
  \begin{equation}
    \prod_{i=1}^T \prob{\sigma_i|\xi_i,T^p} = \prod_{i=1}^t \delta\prs{\sigma_i,\rhd\xi_i},
  \end{equation}
  where $\delta$ is a Kroenecker delta and $\rhd\xi_i=g\prs{S,\xi_i}$. 
  The algorithm $\alpha$ in this case is composed by the description of the activation function
  $g$ \emph{plus} the hypothesis that the information acquisition process, is noiseless.
  
  If we would like to compare, let us say, different theories $T_i$ each corresponding to a
  different student activation function $g_i$, then substitution of the above distribution into
  equation (\ref{equation:Rank}) gives
  \begin{equation}
    \prob{T_i|D_t} \propto \prod_{i=1}^t \delta\prs{\sigma_i,g_i\prs{S,\xi_i}},
  \end{equation}
  where we considered the prior distribution being uniform over all the possible theories. The interesting
  aspect of this equation is that, if any theory misses the correct answer for
  any one of the data points, the deltas guarantee that its probability is reduced to zero. This is obviously a very
  rare case and the reason this happen is that, by considering the theory noiseless, we attributed a strong decisive
  power to the dataset. In this sense, these theories can \emph{in principle} be proved wrong. This means that these
  theories have the main characteristic of being \emph{falsifiable theories}, a concept that about which we will have
  more to discuss in the next section.
\end{example}

\begin{example}[Quantum Mechanics]
  Let us deal with a more familiar example in physics. Consider the theory to be analysed as \emph{quantum mechanics},
  symbolised by $T^{QM}$, and the a question set given by $\Xi=\chs{{\bf x},{\bf p}}$, with ${\bf x}$ the position
  operator and ${\bf p}$ the momentum operator of some system which will be analysed with some experimental setup. In
  fact, the way we are describing the question set is sloppy, as we should actually append a description of the 
  experimental setup. We shall however ignore it for simplicity as this will be no relevant for the point we are trying
  to clarify, as will be seen in the following. 

  In order to simplify the present analysis, we will suppose that the theory describes the system by assuming that
  there is no time evolution between two measurements of these observables. The rules of quantum mechanics state that
  if the system is in the state $\ket{\psi}$ at some instant, then a measurement of an observable $O$ will result in
  one of the eigenvalues $o_i$ of $O$ with probability $|\left<o_i|\psi\right>|^2$ and, after the measurement, the new
  state becomes an eigenstate of that operator. In principle the measurement noise can be reduced to zero as long as
  these rules apply. 

  Let us analyse what happens if the measured dataset is
  \begin{equation}
    D_3=\chs{({\bf x},x_1),({\bf p},p_2),({\bf x},x_3)},
  \end{equation}
  where the first element in the pair indicates the measured observable and the second the
  value obtained. Here we will consider that the data are collected in chronological order 
  from left to right. As the state of the system depends on the result of the former
  measurement, the data is not independent. In fact, the dependence structure can be represented
  by the Bayesian network of figure \ref{figure:BN_QM}, where the state of the system is represented
  by hidden nodes, as it is not measurable and must be summed over. It is important \emph{not to} confuse this kind of
  description with a hidden variable interpretation for quantum mechanics, but just as a graphical way of visualising
  interdependency of each element of its structure.
  \begin{figure}
    \centering
    \includegraphics[width=5cm]{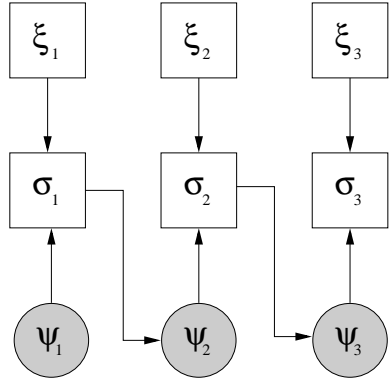}
    \caption{The Bayesian network representing the variables in the quantum mechanics example. The
             dark circles associated to the state of the system represent hidden nodes that must be
             summed over }
    \label{figure:BN_QM}
  \end{figure}
  Then the probability of the dataset 
  is then decomposed, substituting the corresponding values according to the graphical model, as
  \begin{equation}
    \label{equation:QM}
    \begin{split}
      \prob{D_3|T^{QM}} &= \sum_{\psi_1,\psi_2,\psi_3} \prob{x_3|{\bf x},\psi_3,T^{QM}}
                           \prob{\psi_3|p_2,T^{QM}}\\
                        &  \quad\times
                           \prob{p_2|{\bf p},\psi_2,T^{QM}} \prob{\psi_2|x_1,T^{QM}}\\
                        &  \quad\times\prob{x_1|{\bf x},\psi_1,T^{QM}}
                           \prob{\xi_1|T^{QM}}\prob{\xi_2|T^{QM}}\prob{\xi_3|T^{QM}}\\
                        & \quad\times\prob{\psi_1|T^{QM}},
    \end{split}
  \end{equation}
  where the sum over the $\psi_i$'s runs over all possible values of the state of the system
  according to the theory, which is all the possible equivalent classes defining a different
  state in the Hilbert space of the system, and we did not substitute the question in the last factors
  describing the probability of the $\xi$'s for the sake of clarity. In any case, as we are free to choose what to
  measure independent of the order, these will drop out when probabilities are normalised.
  
  As we are considering noise to be negligible, we have that
  \begin{align}
    \prob{\psi_t|\sigma_{t-1},T^{QM}} &= |\left<\sigma_{t-1}|\psi_t\right>|^2,\\
    \prob{\sigma_t|\xi_t,\psi_t,T^{QM}} &= |\left<\sigma_t|\psi_t\right>|^2,
  \end{align}
  where the index $t$ indicates the time step to which
  each variable is related. If we would consider a continuous time index with a time evolution
  between measurements given by a time evolution operator $U$, this would affect the first 
  equation as
  \begin{equation}
    \prob{\psi_{t+\Delta t}|\xi_t,\sigma_t,T^{QM}} =
      |\left<\sigma_t|U^\dagger(\Delta t)|\psi_{t+\Delta t}\right>|^2.
  \end{equation}  
  
  Noise would affect directly equation (\ref{equation:QM}), where we would need to have a 
  sum over the possible answers at each measurement. The probability chain rule would again
  be used as this would be a Markov chain. The first term in the chain would then be written
  as
  \begin{equation}
    \prob{x_3|A({\bf x}),T^{QM}} \prob{A({\bf x})|{\bf x},\psi_3,T^{QM}}
    \prob{\psi_3|A({\bf p}),{\bf p},T^{QM}},
  \end{equation}
  where $A(\cdot)$ corresponds to the answer predicted by the theory at the corresponding
  time step, which are not explicitly written. Let us remind once more that we are
  neglecting a term with the probability of the experiments as their order would not 
  depend on the theory for we are allowed to choose freely if we are going to
  measure either ${\bf x}$ or ${\bf p}$, 
  and they would cancel out with the normalisation. The noise model would
  enter only in the first factor. For instance, white Gaussian noise with unit variance in 
  the measurement process would give
  \begin{equation}
    \prob{x_3|A({\bf x}),T^{QM}} = \frac1{\sqrt{2\pi}} e^{-\frac12 (x_3-A({\bf x}))^2}.
  \end{equation}
   
  In this case, the set of physical constants $\pi$ and the algorithm $\alpha$ are much
  more difficult to describe.
\end{example}
 
For completeness, let us finish again by giving a more formal definition of the process analysed in this section.

\begin{definition}[Modeling]
  The process of creating a theory $T$ that answers the question set $\Xi$ is called \textbf{modeling}.
\end{definition} 
 
\section{Testing}
\label{section:Testing}

Contrarily to what we have done in the last two, we will start this section by defining from the beginning the process
we want to analyse.

\begin{definition}[Testing]
  Given a set of $M$ theories $\chs{T_1,T_2,...,T_M}$ and a dataset $D_t$, the process of calculating the posterior 
  distribution
  \begin{equation}
    \prob{T_i|D_t} = \frac{\prob{D_t|T_i}\prob{T_i}}
                     {\sum_i \prob{D_t|T_i}\prob{T_i}},
  \end{equation}  
  is called \textbf{testing}.
\end{definition} 

Testing in a Bayesian framework then involves the recalculation of the posterior distribution of the candidate theories 
according to Bayes' rule. This can either be done using only the original dataset $D_t$ in order to rank a new theory
(or theories) among the other existent ones or by acquiring new information, thus expanding the dataset to $D_{t'}$,
and again recalculating the rank of the existing theories. Of course, both things can also be done at the same time. 

Ideally, the recalculation of the posterior distributions should be made with the entire
dataset, which in machine learning applications receives the name \emph{off-line learning}.
There may be however situations where old data points become systematically less important or
less reliable. In these cases the posterior can be recalculated for each new data point at time $t+1$ using the posterior at time $t$ as a prior, a practice known as \emph{on-line learning}. Here we will 
deal exclusively with off-line learning situations. References to Bayesian on-line learning can 
be found for instance in reference \cite{Opper98}. 

As we had already shown in equation (\ref{equation:Rank}), discussed briefly in section
\ref{section:Bayes}, in order to calculate the posterior distribution of the candidate
theories given the dataset, a prior distribution over theories should be defined. It is
obviously possible not to favour any theory in the absence of information. However, our experience
tells us that physical theories have some desirable properties. For instance, as we climb up
the ladder that leads to the standard model, the larger the symmetry in the theory. Most of the
efforts related to grand unified theories (GUTs) rely on finding a more encompassing symmetry
group that breaks down to the known gauge groups of electromagnetic, weak and strong
interactions at smaller energies. In this sense, more symmetric theories are preferred to less symmetric ones. 

Deciding about theories that ``explain'' the data equally well is the job of the prior. One of the most used criteria 
in this scenario is the well known principle of \emph{Ockham's razor}, which simply states that if more than one theory 
explains the dataset equally well, the simplest should be preferred. 

Ockham's razor is justified on philosophical grounds based on our belief that
nature ought to be simple, the blame being on us for not being able to describe it 
correctly. ``Simple" is always understood as \emph{mathematically simple}. There is however
no guarantee that nature's laws are indeed simple in these terms. A more rational and practical justification for
Ockham's razor would be that if the theories explain the dataset equally well, we then choose the
simplest one to carry out necessary calculations in order to spend less resources, which now clearly favours theories
with lower \emph{computational complexity}. On the other
hand, machine learning teaches us that in every inference task, there is a trade off between
generalisation ability and fitting (see \cite{Engel01,Bishop06}). By using enough 
adjustable parameters it is eventually possible to perfectly fit any dataset but generalisation,
the capacity of inferring correctly a new data point, becomes compromised.

Whatever the rationale for Ockham's razor is, choosing the simplest theory requires a
precise measure of how complex a theory is, something that we have already discussed in the previous section. 
We will then assume that the prior distribution over a theory $T=\prs{\alpha,\pi}$ must depend on some measure of the 
complexity of the algorithm $\alpha$, which contains the mathematical description of the theory. However, we claim that 
the prior over any physical theory should also depend on the concept of fundamental status. 
As already said, more fundamental theories have less adjustable parameters and more generalisation
power. In order to consider this, note that the prior distribution of any theory can always be written as
\begin{equation}
  \prob{T}=\prob{\alpha,\pi}=\prob{\pi|\alpha}\prob{\alpha}.
\end{equation}

The distribution $\prob{\alpha}$ is the part of the prior that select theories with less 
complexity. The factor $\prob{\pi|\alpha}$ can then be used to select the more fundamental
theory by using some probability distribution that decreases with the cardinality of $\pi$.
In the previous section we argued that more fundamental theories should be more complex. As the exact relation is not
clear at the moment, the above way of writing the prior is general enough to include this dependence in the
term  $\prob{\pi|\alpha}$. In fact, we propose that these are the only characteristics that are needed in the prior. 

As in every Bayesian inference task, the problem of choosing the correct prior is a difficult one, which does not mean 
though that it has no objective solution. For instance, the
\emph{maximum entropy principle} is a method of choosing priors by using Shannon's entropy as
a measure of lack of information and maximising it subjected to constraints that encode all the
information available \cite{Jaynes03}. In this case, if two different observers have the same
amount of information, they agree about their priors. This is no more subjective than the
situation in relativity where two observers may only agree about the size of time intervals or
the order of spacelike separated events if they share the same velocity, although given enough information they can
carry on their own calculations and discover how the other observer is measuring these observables. 
Of course the choosing
of the prior depends on many criteria according to the aims of the task to be done, however the
objectivity means that once these criteria are agreed, there is one way to construct priors
such that again, if two observers have the same information, they end up with the same prior.

A question may arise with respect to the factor of the prior related to the fundamental status of the theory. 
Why not use the
complexity of the set of fundamental constants of the theory as a prior instead of just its cardinality? 
Although this can be done, and it is in some sense the path taken in MDL, in principle fundamental
constants should not be judged by their simplicity and should be learned using the dataset. 
In fact, as we have already discussed when we argued that the whole theory can be reduced to a number to be fed to a 
universal Turing machine, the theory's constants work as a set of indices defining a possibly continuum 
set of theories, each one corresponding to different values of these constants, and
choosing them is equivalent to using equation (\ref{equation:Rank}) to rank the candidate
theories. There are nevertheless situations when it is desirable to compare different
theories, but the exact numerical values of the fundamental constants are not important. The
solution in this case is also the Bayesian one, but now we should sum over the unknown
values of the parameters, which gives the expression 
\begin{equation}
  \begin{split}
    \prob{\alpha|D_t} &= \sum_\pi \prob{\alpha,\pi|D_t}\\
                      &= \sum_\pi \prob{T|D_t},
  \end{split}
\end{equation} 
to which the formula for $\prob{T|D_t}$ can then be applied as before.

A very fundamental idea that must be clearly understood is that we only compare theories constructed with respect to
the same question set $\Xi$. When the question sets of two theories are different, there is no basis to compare them in
general, for they have different scopes. For instance, compare hydrodynamics and quantum mechanics. In principle,
quantum mechanics is able to answer all questions of hydrodynamics (although this assertion has many subtleties), but
it makes no sense to decide between both by comparing their different question sets. Depending on the questions to be
answered, hydrodynamics can even be the preferred theory as we do not need the full power of quantum theory to
calculate flows of water in confined geometries. However, this very example allows us to define a new concept
that relates theories with different questions sets, the notion of \emph{power} of a theory.

\begin{definition}[Power]
  Consider a theory $T_1$ that answers the question set $\Xi^{(1)}$. A theory $T_2$ that
  answers the question set $\Xi^{(2)}$ such that $\Xi^{(2)}\subset\Xi^{(1)}$ is
  said to be \emph{less powerful} than theory $T_2$ and, conversely, $T_1$ is said to be
  \emph{more powerful} than $T_2$. 
\end{definition}

According to this definition, quantum mechanics is more powerful than hydrodynamics, an
assertion that agrees with our intuition for the power of a theory. For the rest of this
work, the word ``powerful" will be used in the strict sense of this definition unless otherwise
stated. In principle, hydrodynamics can be obtained from quantum mechanics by restricting its
question set to a smaller subset relevant to its scope. Another classic example beyond hydrodynamics versus quantum 
mechanics is Newtonian gravity versus general relativity. 
It is clear that for most practical applications the full power of
general relativity is not only unnecessary, but usually undesirable due to the more involved
calculations needed to get an answer as good as the former for all practical purposes. 

Once more, it is not clear what is the correct quantitative measure of power and how a prior over it should be constructed. For instance, a reasonable measure would be the size (or cardinality) of the corresponding question set, 
but nothing prevents us from choosing a monotonic function of these values. This is another point where more research
is required. 

The importance of considering to which question set each theory corresponds when comparing them was already noticed by
Brody \cite{Brody94}. In fact, what we called the question set here corresponds precisely to his concept of 
\emph{scope} of the theory. Brody arrives at pretty much the same conclusions we arrived here that it is immaterial to
compare theories with different scopes, although he failed to accept the Bayesian interpretation, still holding to a pure frequentist interpretation. Again, many of his criticisms find appropriate answers in Jaynes \cite{Jaynes03}.

The ultimate goal of theoretical physics is to find the most powerful theory of all, one that can answer any possible
question set about nature. This is expressed by the idea of \emph{unification}. Unification actually expresses the
sensible human belief that nature is not broken in small non-overlapping domains of unrelated theories, but that it is a
limitation of our body of knowledge to describe it in these terms. The classic example of trying to 
merge general relativity and quantum field theory in a theory of quantum gravity involves this
concept in a subtle way. 

The concept of unification however is subtler than it seems. For instance, if we consider two
different theories answering two disjoint question sets, it may be possible to consider a
larger, more powerful theory than both, that is trivially formed by the union of these two
original theories. By union we mean that all resulting sets should be the union of the specific
sets of each theory. In this sense, disjoint question sets should be understood as sets such
that the sets formed by their answers are independent. Although in this trivial case we would not be too compelled to
call the resulting theory a true unified theory, we can use it as a guide to define more formally this concept.

\begin{definition}[Unified Theory]
  Consider any two question sets $\Xi^{(1)}$ and $\Xi^{(2)}$. A \emph{unified theory}
  is a theory that can answer the union of these sets $\Xi^U=\Xi^{(1)}\bigcup\Xi^{(2)}$.
\end{definition}

Note that a unified theory is always, by definition, more powerful than any theory answering only $\Xi^{(1)}$ or
$\Xi^{(2)}$, but there is nothing that guarantees that it is more fundamental. In fact, the trivial unification based on the union of two theories is actually less fundamental as it will always have at least the same number of constants
as the individual theories that it is composed of. 

Returning to the example of quantum gravity, let us consider the approaches of Loop Quantum
Gravity (LQG) and String Theory (ST). The fundamental problem is that there is a question 
set $\Xi^G$ which is answered by general relativity in a highly successful way. But as gravity is 
supposed to be universal and quantum systems also gravitate, 
their question sets overlap at some point and these theories should be unified somehow. 
The first attempts however have resulted only in theories that were not
able to be made consistent. Therefore, both LQG and ST try to be unified theories in the sense
that they both try to answer the union of two different question sets. However ST aims to be
a more fundamental theory as it actually is argued to rely on less adjustable parameters, 
assuming the existence of less fundamental physical constants.
This can be seen by comparing both to the standard model (SM). LQG is not more fundamental
than the SM as it has at least the same amount of adjustable parameters corresponding to masses,
couplings and mixing angles for, being a theory just about gravity, it does not change the
basic set up of the SM. On the contrary, ST has less parameters than the SM in the sense
that many of the physical constants that must be obtained from experiments in the latter are 
in principle calculable in the former without the need to be learned from any dataset. 

Quantum gravity is also a good example to understand how difficult it is to assess the
complexity of a theory in general. The catch is that it is not a fully developed theory. 
Even if either ST or LQG, or maybe both, are correct, many calculations cannot be done yet in
these framework for they are too complicated and not completely understood. This means that
there is not yet, for instance, an algorithm $\alpha^{ST}$ that allows ST to give
answers to the full question set it is supposed to answer and, therefore, no way to estimate
its full complexity in order to compare with other possible theories, although partial 
comparisons using restricted question sets can be done. However, many theoreticians believe
that fundamental status is an even more desirable characteristic than simplicity and use the fact
that ST is arguably more fundamental in its favor.

\section{Falsifiability}
\label{section:Falsifiability}

Falsifiability is such a delicate and fundamental concept related to physical theories that
it deserves a separate discussion. String theory, for instance, is usually attacked on the
grounds that it may not be falsifiable. The main difference between science and religion also
is supposed to rely on falsifiability. 

Falsifiability is a concept related to the testing process of a theory. It is well known that not being falsifiable does not
mean that a theory does not agree with the experiments. For instance, the trivial case of a theory that has no axioms,
only adjustable parameters corresponding to the probabilities of all possible answers to all questions cannot be ever
disproved. 

Consider, once more, the case of a polynomial of degree $n$. Any dataset with $n+1$ or less experimental data points
obtained from some function can be fitted by this polynomial. However,
one more data point can be enough to test if the polynomial is or is not the desired function.
In the more general case of inferring probabilities for answers (what we called the theoretical answers) in order to be falsifiable, a
theory must be able to answer some question in such a way that some new data would be able to disprove it, where by 
``disproving" we mean that the probability of that theory given the 
data would be decreased to zero after this new dataset is measured. 
In terms of the formalism developed up to this point, we will explore a possible definition of falsifiability.

According to equation (\ref{equation:Rank}), the only way to reduce the posterior of a theory
to zero is if the probability of the dataset is zero given the theory, equivalently, if the theoretical answer evaluated for that dataset is identically zero. If there is one dataset
for which this happens, then the theory can be considered to make a prediction which is falsifiable if experiments can be tailored to check the theory's predictions about this dataset. Of course this requirement is null if there is only
one dataset to consider. The concept of falsifiability depends on the existence of new data to be measured, or new
information to be acquired. This suggests the following definition.

\begin{definition}[Falsifiability]
  A theory $T$, answering a question set $\Xi$, is called \textbf{falsifiable} if there exists more data points to be 
  obtained beyond those used to construct the theory and at least one dataset $D_t$ with zero probability
  \begin{equation}
    \prob{D_t|T} = 0.
  \end{equation} 
\end{definition}

The first observation is again the strong dependence of the definition on the considered
question set. A falsifiable theory can give origin to a non-falsifiable one if the question
set is restricted to questions such that $\prob{D_t|T}\neq0$ for any possible dataset. 
This implies that for any theory, a concept of
absolute falsifiability could therefore only be defined if the set of all addressable questions is known. 
If this set is not fully known, it is not possible to claim that a theory is not
falsifiable in principle, although it may not be falsifiable in practice in the sense that the
theory restricted to the part of the question set that is experimentally accessible may not be.
 
This leads to the somewhat obvious conclusion that a theory cannot be said to be absolutely non-falsifiable until all the question set it can answer is known. By this definition, for
instance, there would be still no grounds at the moment to say that ST is fundamentally non-falsifiable. In fact, many 
theories of quantum gravity are in this position as they deal
with high energy phenomena that are inaccessible to our present accelerators. 
Another example would be Hawking radiation. It is a falsifiable prediction once it depends
in principle only on more advanced technology to be measured, although it is not possible to
do it at the present. 

The second point to be noted is that, if there is no more data points to be taken, there is no way to falsify the theory.
This looks like an obvious assertion and, indeed, it is. However, it is probably false that a situation like this
will ever happen in physics. Although a Theory of Everything is supposed to account for every possible phenomenon, still
there is no way to claim that all possible measurements will ever be done. A possible, hand waving argument, would be
to use the holographic principle to argue that, as observations seem to imply that the universe will never stop
its expansion, if we assume that the information contained in the universe is given by the size of its horizon, this 
information will be ever increasing. This would mean that even if at some time $t$ all the information about the 
universe was collected in the dataset $D_t$, at $t+ dt$ more information would be created and so on \emph{ad infinitum}.

Although falsifiability is a difficult criterion to evaluate in a theory, in the previous
section we saw that the more fundamental constants (adjustable parameters) a theory has, the
easier it is to agree with the data. Therefore it is reasonable to expect that the less
fundamental a theory is, the smaller the chances that it is falsifiable. According to our claim that more
fundamental theories should be more complex, we would than be led to the conclusion that a less complex theory is not
necessarily more falsifiable. The ``everything goes" theory for instance has only this axiom
and an infinite number of adjustable parameters to be measured by experiments, being fundamentally non-falsifiable, which agrees with our claim. 

Another kind of prediction that a theory is usually expected to make is to derive and answer new questions previously 
not contained in the original question set $\Xi$ used as a basis for the construction of the theory. 
There are many arguments in favour of requiring that a theory should be able to do so. The practical one, for instance, 
is that this kind of prediction is the one that leads to
technological advancements, which leads to an enhancement of society's quality of life. It also
brings economic wealth. This concept will be called the \emph{predictive power} of the theory.
The trivial ``everything goes'' theory stated above is not only non-falsifiable, but also has no
predictive power at all.

Note that predictive power and falsifiability are related, but are different concepts. A theory
can make predictions beyond the original dataset that may not be falsifiable and a theory may
be falsifiable but do not address questions beyond the original question set.

However, again, intuition says that the larger the predictive power of a theory, the more
probable it is to be falsifiable. This can be understood if we note that the higher the 
predictive power of a theory, the larger becomes the question set and, therefore, the larger
is the number of possible datasets, increasing the probability that at least one is 
impossible to be observed. 

This culminates in the two most important definitions of this paper: \emph{scientific theory} and \emph{scientific method}. Although these definitions were actually stated before, unlike the situation up to now, each one of the concepts involved is now precisely defined through the sequence of all previous definitions in this work.

\begin{definition}[Scientific Theory]
  A theory $T$ is called a \textbf{scientific theory} if it is falsifiable.
\end{definition}

\begin{definition}[Scientific Method]
  The iteration of the processes of acquiring information, modeling and testing in any order necessary to produce
  and rank \emph{scientific} theories that answers a specific question set $\Xi$ is called the \textbf{scientific
  method}.
\end{definition}

We require that \emph{all physical theories should be scientific theories} unless all the information available in
the universe entire chronological history is contained in a hypothetical obtained dataset $D_\infty$. As discussed 
above, the existence of this dataset may even be impossible \emph{in principle}, although we are far from a rigorous
proof. In any case, if all possible experiments were really done and its results cataloged, science would
be truly over and the above questions would be simply meaningless, the only work really remaining being a better way of 
compressing the information in the most mnemonical way.

\section{Cosmology}
\label{section:Cosmology}

As the first cosmological example to be analysed in this work, we consider an interesting question which has been revived recently and is related to the concept of \emph{typicality} in cosmology. Bayesian selection of physical theories has been conjured in a number of papers \cite{Page07,Hartle07,Page08} to justify or argue against this concept. Let us examine these ideas using some of the framework developed here.

The idea that theories where our observations are typical are favoured against others where they are not is expressed more specifically in
\cite{Page08}. The argument boils down to the fact that, in the case of equal priors, the probability of the theory
depends only on the dataset
\begin{equation}
  \prob{T|D_t} \propto \prob{D_t|T},
\end{equation}
and then the theory that predicts that the data has a higher probability, i.e. is ``more typical'', is preferred.

Of course the line of reasoning could not be more clear. Indeed the best theory given equal priors gives the higher
probability to the dataset. Actually, the best one is the one which gives the dataset probability 1 in expense of 
other possible datasets. If the dataset $D_t$ contains all data that will ever be possible to measure, then that is
the only possible choice. But let us analyse better the role of the dataset.

First, let us define the question set. Just one experiment is enough, namely $Q=\text{'What kind of observer?'}$. Let
us allow for $N+1$ possible answers, i.e., $\mathcal{A}=\chs{0,1,...,N}$ with 0 corresponding to no observer, 1 to a 
human observer and the other integers to different kind of observers like aliens, Boltzmann brains, etc. The 0
value is necessary here because the experiment is designed to \emph{find} an observer somewhere. If no observer is found, then the zero value must be attributed. For example, we can divide the visible universe into a grid and attribute to each region of the grid one of these numbers corresponding to the result of the experiment $Q$. Of course, this grid can be then made as thinner as we want up to the size where physical principles will affect our experiment.

As we have already described, we could sum over the regions to answer the question of ``what kind of observer?''  can be found anywhere, but
the experiment should still be well defined and the probability of finding some observer somewhere must be provided by
the theory even if we marginalise over it at the end. By defining then all the necessary elements, 
we can discuss again the typicality issue.

Suppose now that the only place we looked for is on the Earth. Most of our data will have the value 1. The theory that says
that humans are the only type of observers would be the best. As long as we have other places to look at, this changes.
Actually, as far as we know, ``no observer" is the typical result for the universe. Clearly this is a simplification, 
but it is not far from the ones used in the above papers. The most important lesson is how decisive is to define well 
the question set.

The second issue we are concerned with is related to the multiverse hypothesis. Consider the idea contained in \cite{Page07}, the suggestion that multiverse theories are always 
favoured by Bayesian reasoning for they predict probability 1 of existence for 
any value of a measurement while single universe theories do not, implying that the likelihoods of the former are 
always larger than of the latter for the existence of these values. Note that the multiverse theory gives probability one for the ``existence" of solutions for some question set $\Xi$ while the universe gives less than one. Therefore, assigning equal 
priors for both, the likelihoods will select the multiverse theory. In order to analyse this 
assertion, there is again the need for a precise definition of what are the questions the theories must
answer. In this case the question set is not the original $\Xi$, but a derived one $\Xi'$ that contains
the unique question $Q'=\text{'Does the set of answers for $\Xi$ exists?'}$ The possible range of
the answers for this question is the set $\mathcal{A}'=\chs{\text{Yes},\text{No}}$. By definition,
the probability distribution of the answer $A(Q')$ in the multiverse theory $M$ is given by
\begin{equation}
  \prob{A(Q')|Q',M}=\delta\prs{\text{Yes},A(Q')},
\end{equation}
which means that the probability of a 'Yes' is one and of a 'No' is zero. The same distribution
for the universe theory $U$ is
\begin{equation}
  \prob{A(Q')|Q',U}=P\delta\prs{\text{Yes},A(Q')}+(1-P)\delta\prs{\text{No},A(Q')},
\end{equation}
which gives probability $P$ to 'Yes' and $1-P$ to 'No'.

Once more, comparing the theories given some data is the desirable scenario. If the answer is
'No', then obviously the multiverse is ruled out, so let us suppose that our data says that the answer is 'Yes'. There 
are three cases:
\bigskip

(1) $P=1$. Then both theories give the same likelihood for the data. If they have equal 
priors, both theories have the same probability and cannot be decided on the basis of the
dataset. 
\bigskip

(2) $0<P<1$. Then the likelihood of the data given by the multiverse is really higher and the
multiverse solution must be preferred. Although this may sound strange, the point is that we
are now certain that the data exist and the universe theory predicts that the data may not, so
it is, based only on the data, obviously less desirable 
\emph{if these are the only two alternatives}.   
\bigskip

(3) $P=0$. The universe is obviously ruled out as it gives the wrong answer.
\bigskip

Note that in this example, the nature of the question admits only a precise definition. The 
distributions of the answers are Kroenecker deltas, which means that they have no spread. Like
in the previous case, there is not a definite answer that is correct always. The 
\emph{posterior} probabilities again depend on the data and can be defined only by it.

The specific way in which the likelihoods are constructed shows that the advantage of multiverse theories over universe 
ones is not straightforward. The above defined ``theories'' are perfectly fine from the formal point of view.
However, different requirements must be taken into consideration when constructing the theories. As already
discussed, there are many of them. If each element is not well defined there is not a proved superiority of one over 
another in the present search for a theory that
describes our world. This superiority as in any other physical theory must be decided on the
basis of collected data which, at the present, is not enough for a decision to be made.

Finally, as the last cosmological issue to be analysed, let us consider the weak version of the anthropic principle, as 
the strong one is non-falsifiable. The analysis will be very simple as we only want to make the point that this 
``principle'' is only a label for a noiseless observation in a broader dataset. The noiseless observation is the one
that humans exist. The broader dataset is the data accumulated about the physical and chemical requirements that are 
necessary for us to live. Given these requirements, and the fact that we indeed exist, any theory related to \emph{any}
question set to which this is relevant, should have answers with distributions that give reasonable values for these
datasets.

The fact that the anthropic principle can be used to calculate the range of many physical constants can simply be 
reformulated according to the framework developed in this paper as the assertion that the theories that do not take
into consideration this \emph{noiseless} information from the beginning have automatically zero probability. The fact
that the constants calculated in this way may be not specific values but be in some range just reflect the fact that 
the other observations about what is needed for \emph{human} life is not known in a noiseless way. From this point of 
view, the importance status given by the word ``principle'' becomes highly questionable.

\section{The Isolated Worlds Problem}
\label{section:Problem}

There is no physical principle presently known forbidding the existence of some parts of our universe that never interacted and will never do so among themselves, although this can be seen as a highly speculative idea. This idea is the basis of what we will call the \emph{isolated worlds} problem, which can be described in the following way. Suppose that there are two (or eventually more) regions of the universe which never interacted and will never do so by definition. Apart from that, both regions have physical laws
that allowed the development of intelligent life. Each one of these regions will be called an \emph{isolated world}. 
According to the usual scientific considerations, as there is nothing from one
region that can be measurable by observers from the other, each region should be considered by the observers living in the other as non-existing. However, by our hypothesis, both regions do exist. Is there a fundamental problem here? Is science unable to address this question? Or is this question actually non-sensical? 

As already argued, the fact that no physical principle forbids the described situation renders the argument that this is a non-sensical question void. We will actually argue that this question is indeed addressable, in principle, by means of our framework if a special condition is met. Suppose that there are many theories available to describe our universe and, when compared to the
available datasets corresponding to all the physical knowledge acquired up to that moment, one of them is ranked with
a probability much higher than any other. Now suppose that the mathematical structure of this theory not only predict the existence of two isolated worlds according to the above definitions, but that in fact requires it as an integral part.

Now, even if there is no way for an observer living in one of the isolated worlds to measure directly anything from the other, the whole framework developed here forces us to admit that the probability of both worlds existing should be considered higher than that of both not existing and, if the addition of more datasets increases
the probability of this specific theory, the more sensible is to considered the existence of both as real. A more surprising conclusion is that, if the theory requiring them is falsifiable, we should assume that the existence of these two isolated worlds is also falsifiable. 

It is possible to argue that the above discussion is based on highly questionable speculations. We can counter-argue that by saying that, although speculative, none of the arguments above has any physical impossibility of being correct, meaning that there is no no-go theorems forbidding them. However, more strong support can be found in the fact that there is indeed a cosmological theory where a similar situation exists: \emph{eternal inflation} \cite{Guth01}. In the eternal inflation scenario, new universes are being continuously formed, and these universes do not interact at all.

This interesting analysis shows that the Bayesian framework outlined here is capable of extending the scope of scientific inquiry to questions beyond what has been usually accepted as addressable.

\section{Conclusions}
\label{section:Conclusions}

The main objective of this work was to develop a formalisation of the scientific method using the framework of Bayesian inference. We used this framework to describe the basic processes that compose the
scientific methodology. By relying on this structure, we were able to give formal definitions to many important, albeit previously only intuitive, concepts that are used to characterise theories of physics. Many of these concepts lack a clear boundary between themselves. The proposed formalisation allowed a sharper definition of this boundary and a more precise study of them, providing a better understanding of their relevance in the structure of physical theories. The sequence of definitions presented culminated in the two most important ones in the paper, those of a \emph{scientific theory} and of the \emph{scientific method}, both agreeing with all our intuitive requirements.

Many important conclusions can then be drawn by the use of the formal structure developed in this work. The first important insight is that any theory is only defined with respect to some set of questions and their precise definition is absolutely crucial for
comparing two or more theories. This implies that comparison between theories that answer different question sets is
not a sensible procedure in general. The question sets define the \emph{scope} of the theories to be compared, which
do not need to be theories about everything but can be restricted to a limited number of questions of interest to the
researcher of some scientific area or simply some specific system. This allows the application of the methods presented here to select the most appropriate theory to describe that limited set of
questions which, as already argued in the main text, does not need to be the most powerful or fundamental theory in 
general. 

Another very important insight obtained from this work is the role played by the theory's fundamental constants, which in physical theories are actually what is meant by their 
fundamental \emph{physical constants}. As a consequence of this identification, it was possible to make a clear
distinction between these constants $\pi$ and theory's algorithm $\alpha$, which then allowed the formalisation of the
concept of fundamental status of a theory, which is in full agreement with the term as used in theoretical physics 
today. The more difficult question of the complexity of a theory then boils down to the analysis of 
the complexity of the algorithm $\alpha$. We have discussed the difficulties in defining a unique measure of complexity, with
many attempts to do so in the literature. Still, based on our studies, we argued that it is sensible to conjecture that \emph{the more fundamental a theory is, the more complex it tends to become} although we still
cannot prove this assertion rigorously as a theorem, as there is still a lacking of agreement about what should be meant rigorously by this complexity. Still, this conjecture can be seen as an instance of the well known and thoroughly studied issue of the interplay between generalisation ability and fitting in machine learning, and its content is supported by the results in this area.

Note that the inclusion of the \emph{noise term} in the \emph{modeling} of the theory instead of in its \emph{testing} is an incorporation of the ideas expressed by Duhem \cite{Duhem14} that the interpretation of experiments is theory-dependent. Although Duhem do not recognise \emph{simplicity} as a valid criterion to choose one theory among others as we considered in this work, this does not invalidate our analysis. The reason for this is that simplicity in our work is an \emph{a priori} requirement that enters in the construction of the prior distribution. Once simplicity is incorporated to it, this concept is not used again during inference process itself.

Another key issue is that it becomes clear that although Bayes' rule is a fundamental principle of theory selection,
and therefore of the scientific method, it is not sufficient to capture all the concepts of importance to science. Other elements,
like falsifiability and fundamental status, which are not directly related to inductive reasoning are also important. One possibility is that these principles are still derivable from a more fundamental principle. In our opinion, which is based on the general feeling of the Bayesian community, this principle is probably \emph{maximum entropy}. One indication that this may be the correct path to take is the fact that even Bayes's rule seems to be derivable from it.

By applying the developed framework to some cosmological problems, we arrived at the following conclusions. With respect to the typicality of the human observer, we saw that the exact formulation of the question is of utmost 
importance and no proper answer can be given without a better posed question. We concluded that there is no basis for 
favouring multiverse theories on the proposed theoretical basis only and any preference can only be attributed to experimental data.
Finally, we argued that attributing the status of a physical principle to anthropic reasoning is not appropriate as
it can be viewed as a very simple case of inferring the structure of a physical theory from two sources: (i) the very trivial observation (or datapoint) that we exist, (ii) the posterior probability resulting from all experimental available data (collected literally through thousands of years) about the conditions that we need for our survival.

The last result of this work is another philosophical important one. We showed, through means of a problem we called the \emph{isolated worlds} problem, that the framework we developed can show that through pure inference, science can address situations where a strictly positivist approach would render it useless and, above that, would not be considered a valid scientific question. This shows that using Bayesian inference, the extent to which science can be used is enlarged to situations which were beyond it in other formulations. 

Although we did not discussed the concept of truth in the whole paper, as we indeed said we would not, we need to
include a comment about it. The representation of the probabilistic dependencies of a theory, which we argued 
can be expressed in the form of a Bayesian network, allows for the introduction in this same theory of hidden nodes. When some variable appears in a theory \emph{only} as a hidden node, it is fair to discuss on philosophical grounds if any reality can be attributed to this variable. The positivist viewpoint answers this question as a ``no'', while the 
mathematical universe hypothesis, on the contrary, would answer it as a ``yes''. 

Concerning the differences between our framework and Minimum Description Length \cite{Li97}, MDL suggests to choose theories by minimising the complexity of the description of the hypothesis plus the dataset. Although MDL may be desirable for many applications and a good 
approximation to Bayesian inference, our framework allows us to
address questions that do not appear in MDL. These questions, like falsifiability and
unification, play a very important role in the development and selection of physical theories
and we hope that the use of our formalism can lead to a better understanding of them. 

As a final comment, let us highlight that by using ideas coming from probability theory and machine learning we were
able to give a mathematical framework to questions that could be considered to lie only on the sphere of philosophy. This shows how important is the role played by pure philosophy in the scientific endeavour, something that seems to be forgotten nowadays.

\begin{acknowledgements}
I would like to thank Dr. Juan P. Neirotti and Prof. David Saad for stimulating discussions. The comments and suggestions made by Prof. Ariel Caticha about the ideas in the manuscript were deeply inspiring and thoughtful. I would like to thank him for taking the time to read this work and apologise for those points where I did not follow his suggestions, for what I should take alone all the responsibility. I also would like to thank Prof. Nestor Caticha from the University of Sao Paulo, where the main part of this work was done, for introducing me to Bayesian theory and helping me to see its relevance not only to physics, but to science as a whole.
\end{acknowledgements}

\bibliographystyle{spphys}       
\bibliography{master}   

\end{document}